\documentclass[]{aa}
\usepackage[dvips]{graphicx}
\usepackage[dvips]{graphicx}
\usepackage{psfig}
\newcommand {\chem}[2] {$\rm{}^{#2}\kern-0.8pt#1$}
\begin{document}
\title{Modelling of Intermediate Age Stellar Populations: \\
III- Effects of Dust-Shells around AGB Stars}
\author{M. Mouhcine }
\institute{Observatoire Astronomique, Universit\'e L.\,Pasteur \& CNRS: 
UMR 7550, 11 rue de l'Universit\'e, F-67000 Strasbourg}
\date{Received ; accepted ..}
\titlerunning{Dusty AGB Stars in Intermediate-Age Populations}

\abstract{
In this paper we present single stellar population models of intermediate 
age stellar populations where dust-enshrouded Asymptotic Giant Branch (AGB) 
stars are introduced. The formation of carbon stars is also accounted for,
and is taken to be a function of both initial stellar mass and metallicity.
The effects of the dusty envelopes around AGB stars on the 
optical/near-infrared spectral energy 
distribution were introduced using semi-empirical models where the mass-loss 
and the photospheric chemistry determine the spectral properties of a star
along the AGB sequence. The spectral dichotomy between oxygen-rich stars and 
carbon stars is taken into account in the modelling. \\
We have investigated the AGB sequence morphology in the near-infrared 
colour-magnitude diagram as a function of time and metallicity. We show that 
this diagram is characterized by three morphological features, occupied 
by optically bright oxygen-rich stars, optically bright carbon stars, 
and dust-enshrouded oxygen rich and carbon stars respectively. Our models 
are able to reproduce the distribution of the three AGB subtype stellar 
populations in colour-colour diagrams. Effects of dusty envelopes on 
the luminosity function are also investigated.\\
We have extended our investigations to the integrated spectro-photometric
properties of stellar populations. We find that the contribution of AGB 
stars to the near-infrared integrated light decreases, making 
optical/near-infrared colours of intermediate age populations bluer than 
what is expected from pure photospheric emission models.
%%%%%%%%%%%%
}
\maketitle
%%%%%%%%%%%%%%%%%%%%%%%%%%%%%%%%
%--------------------------
\section{Introduction}
Stars in the AGB phase are a dominant source of the near-infrared light of 
intermediate-age (0.1\,Gyr$\,\le\,$age$\,\le\,2\,$Gyr) stellar populations. 
As an example, AGB stars (defined as stars with M$_{bol}\,\le\,-3.6$) 
contribute up to 60\% of K-band light at 1\,Gyr (Ferraro et al. 1995, 
Mouhcine \& Lan\c{c}on 2002a).

AGB stars develop at the end of their life a strong mass loss. The optical 
light of AGB stars with the highest mass loss rate is almost entirely absorbed 
by their dusty circumstellar envelopes and re-emitted at longer wavelengths.
These obscured objects become very bright infrared sources with particularly 
red near-infrared colours. Omont et al. (1999) have shown that at least 25\% 
of the RGB stars in the galactic bulge are losing mass, based on their very 
red ISO colours. 

Another characteristic of AGB stars is the formation of carbon stars 
(i.e., AGB stars with C/O$\,>\,1$). It has been well known from counts 
in the Magellanic Cloud clusters (e.g. Aaronson \& Mould 1980, 
Frogel et al. 1990, Rebeirot et al. 1993) that carbon stars represent 
a significant fraction of the AGB population of intermediate age clusters, 
at least at sub-solar metallicity. Hence, they are responsible of a 
significant fraction of the integrated light of intermediate age stellar 
populations (Costa \& Frogel 1996).
In addition, carbon stars tend to be the intrinsically brightest of the AGB 
stars present, as they usually correspond to later evolutionary stages than 
their M-type counterparts. It is well established observationally, at least 
for simple stellar populations, that carbon stars are redder than M stars 
(Persson et al. 1983).

Newly large and homogeneous statistical sample of photometric data sets in 
the near-infrared (2MASS, Skrutskie 1998, and DENIS, Cioni et al 1997) and 
in the mid-infrared (ISOGAL, Omont et al 1999), and in the optical 
(Zaritsky et al. 1997) give us a unique opportunity to study the late-type 
stellar content of our galaxy and the Magellanic Clouds, and to draw 
conclusions about their formation history.

The observed structure of the Hertzprung-Russell (HR) diagram of resolved 
galaxies is a convolution of two functions. 
One function describes the rate of the formation of the stars which are 
present today in a galaxy. The second function describes the evolution 
of a star in the observed HR diagram (i.e., lifetime of different stellar 
phases, the location of those phases in the HR diagram). At identical 
star formation histories, the apparent 
morphology is controlled mostly by the second function. Accurate stellar 
evolutionary models that cover the necessary range of ages, metallicities 
and evolutionary phases are needed to retrieve information about the 
star formation history of a galaxy. Unfortunately, the actual state of 
our understanding of late-type stellar evolution is still far from being
reliable enough to perform such a goal accurately. The work presented in 
this series of papers aims at improving our knowledge of the effect of late 
type stars on stellar population properties.

Effects of the dusty envelopes surrounding AGB stars, in addition to those 
related to the formation of carbon stars, on (i) the morphology of the AGB 
sequence in the observational HR diagram of resolved 
stellar populations and (ii) integrated properties of unresolved populations 
have been neglected for a long time. Mouhcine \& Lan\c{c}on (2002a) have 
presented a new spectral library of single stellar populations where the 
formation of carbon stars was taken into account. They showed that the 
presence of carbon stars in the stellar populations lead to redder integrated
near-infrared colours. Bressan et al (1998) have presented single burst 
population models taking into account the effect of the formation of dusty 
envelopes associated with AGB mass loss. They have used a radiative transfer 
model to correct the stellar spectra predicted by standard evolutionary 
tracks, assuming that all circumstellar envelopes have the same chemistry
(i.e., silicate or carbonaceous grains). Hence when they consider carbon 
rich dust, the whole dust-enshrouded stellar population have the same 
properties, even low core mass stars that never form carbon stars. 
In addition, they have not taken into account the spectral dichotomy 
between oxygen rich and carbon stars, considering giant star spectra 
to be representative of the spectral energy distribution of AGB stars.  

Based on the models of Mouhcine \& Lan\c{c}on (2002a), we constructed 
a set of theoretical models which account simultaneously for the formation 
of carbon stars and the dust shells around thermally pulsating-AGB 
(TP-AGB hereafter) stars, and obtained a new set of isochrones suited 
for the analysis of the near-infrared data. Our main goal, as noted above, 
is to study the evolution of the AGB sequence morphology and integrated 
near-infrared properties.

In Sect.\,\ref{syn} we present the grid of stellar evolutionary tracks 
which we use in order to calculate isochrones in the theoretical plane 
(i.e. luminosity vs. effective temperature). We recall briefly the modelling 
of the TP-AGB evolutionary phase, which is included by means of a synthetic 
model including all relevant physical processes that control the evolutionary 
properties that are related to their effects on integrated properties. 

We describe the dusty envelope model used to calculate the effects of 
circumstellar shells around late-type stars on the energy distribution, 
and calculate the isochrones in the observational plane 
(i.e., magnitude vs. colour).

In Sect.\,\ref{pro} we study the effects of various TP-AGB subtypes on 
the evolution  of near-infrared properties of both resolved and unresolved 
stellar populations. We discuss how they depend on age and metallicity. 
Finally, in Sect.\,\ref{con} our conclusions are drawn.

\section{Population synthesis modelling}
\label{syn}
\subsection{AGB stellar evolution}
Synthetic evolution models represent an attractive tool to investigate the
evolutionary and chemically aspects of AGB stars. The evolution of TP-AGB 
properties are followed by means of such models. 
We present briefly the stellar evolution models that we used to follow
the formation and the evolution of TP-AGB stars. The reader is referred
to Mouhcine \& Lan\c{c}on (2002a) for more details.

The model includes different physical processes affecting the evolution of 
AGB stars along the TP-AGB phase and playing a dominant role in determining 
the lifetime, the extent of nuclear processing and the chemical abundances, 
like the envelope burning, the third dredge-up, the high rate mass loss at 
the tip of the TP-AGB phase. 
We start from the last model calculated along the early-AGB evolutionary 
tracks. The total mass, core mass, effective temperature, bolometric 
luminosity of the models, and carbon to oxygen ratio in the stellar 
envelope are let to evolve according to semi-analytical prescriptions. 

Synthetic modelling of the TP-AGB phase requires the 
knowledge of: (i) the critical core mass M$_{c}^{min}$ above which 
dredge-up is triggered, (ii) the dredge-up efficiency (defined 
as $\lambda=\Delta M_{dredge}/\Delta M_{c}$, where $\Delta M_{dredge}$ 
and $\Delta M_{c}$ are, respectively, the amount of mass dredged-up 
to the envelope after a pulse and the increase in core mass during the 
time between two successive TPs), (iii) the critical envelope mass above 
which the envelope burning is effective, and (iv) the element abundances 
in the inter-shell material. 

The basic physical prescriptions which the models stand on are the 
following:
\begin{enumerate}
\item the core mass/luminosity relation; 
\item the core mass/interpulse-period relation;
\item the core mass evolutionary rate;
\item an algorithm to calculate the effective temperature;
\item CNO burning possibly occurring in the convective envelope base;
\item prescription to calculate the mass loss rate by stellar wind as 
      function of the stellar parameters;
\item the third dredge-up and its efficiency;
\item the composition of the third dredge-up material.
\end{enumerate}

The prescriptions used account for different physical processes 
affecting the core mass/luminosity and the core mass/interpulse-period 
relations, and do not rely on the assumption that the classical linear 
relations are valid. 
Marigo et al. (1999) have shown that the actual state of semi-analytical 
modelling account for both breaks of the core mass-luminosity relation, 
for low-mass stars when diffusive overshooting is introduced (Herwig et 
al. 1998), and for massive stars due to the envelope burning (Bl\"ocker
1995). The adopted relations are rather good representations of the 
outputs of the detailed numerical models (Wagenhuber \& Groenewegen 1998).

The theoretical values of the dredge-up parameters are uncertain. Standard 
numerical models predict low dredge-up efficiencies (Boothroyd \& Sackmann 
1988), while models assuming diffusive overshooting predict highly efficient
dredge-up events (Herwig et al. 1998). 
Actually both parameters $\lambda$ and M$_{c}^{min}$ affect the lifetime 
of carbon-rich phase and the age interval for which carbon stars are present 
in a stellar population. Both parameters are also expected to depend on 
properties such as the stellar metallicity or its instantaneous mass and 
structure, but the available information on these relations is neither
complete nor easy to extrapolate (e.g. Herwig 2000, Marigo et al. 1999).
Most of stellar evolution calculations fail to predict carbon stars 
at luminosities as low as observed, and are only able to obtain dredge-up 
for relatively high initial stellar masses (Forestini \& Charbonnel 1997).
The models presented here assume constant $\lambda$ and $M_c^{min}$. With 
this assumption, reproducing the luminosity function of carbon stars in 
the LMC requires intermediate dredge-up efficiencies and leads us to adopt 
$\lambda=0.75$ and $M_c=0.58$ (Groenewegen \& de Jong 1993, Marigo et al. 
1996). \\
The physical conditions needed to trigger the envelope burning, 
and its efficiency, are still uncertain and matter of debate. We assume 
that the envelope burning is effective only if the envelope mass exceeds
a critical value (see Marigo, 1998, for alternative approach). To estimate
this critical value, we use the analytical relation presented by Wagenhuber 
\& Groenewegen (1998). \\
The chemical composition of the dredged-up material is taken from 
Boothroyd \& Sackmann 1988.

\begin{figure}[htbp]
\includegraphics[clip=,angle=0,width=9.cm,height=9.cm]{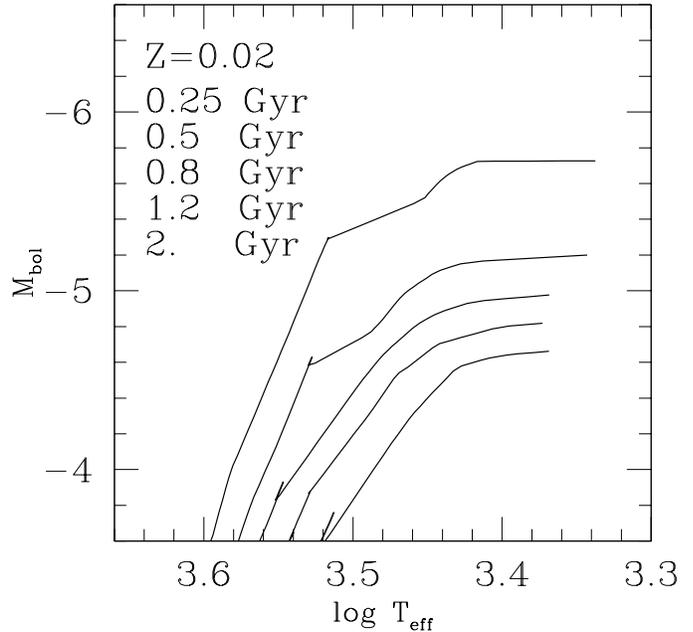}
\caption{Isochrones in the cool luminous region of the HR diagram 
populated by AGB stars for the age interval when these stars dominate 
the near-infrared light of stellar populations. The isochrones are 
calculated for [Z=0.02, Y=0.28] initial chemical composition. 
The younger the isochrones, the brighter they are. Adopted stellar 
evolution free parameters are discussed in the text.}
\label{theo_hr}
\end{figure}

The final calculations are performed using a mixing length parameter 
$\alpha\,=\,2$, Bl\"ocker's (1995) mass loss prescription with a mass 
loss efficiency of $\eta\,=\,0.1$ (Groenewegen \& de Jong 1994). 
The end of TP-AGB phase is determined by the ejection of the stellar 
envelope under the influence of the stellar wind.
The models provide the lifetime of the whole TP-AGB phase and its subtypes 
phases, in addition to the effective temperature and luminosity, as functions 
of time and initial mass. All this information is used to determine the 
spectral type of each star in the HR diagram. A crucial key that determines 
the nature of intermediate age populations, is the sensitivity of the TP-AGB 
star properties to the initial metal content. This will affect in particular 
the formation of carbon stars. We would like to note that the models discussed 
here have shown their ability to reproduce relevant constraints relative to 
intermediate age stellar populations as observed in the solar neighbourhood, 
in Magellanic Cloud globular clusters, or in Local Group galaxies (Mouhcine 
\& Lan\c{c}on 2002\,a,b). Fig.\,\ref{theo_hr} shows the predicted AGB sequence 
in the HR diagram for the stellar populations dominated by AGB stars at solar 
metallicity.

Let us discuss first what these models tell us about the effects of carbon 
stars and dust-enshrouded stars on the properties of single stellar populations. 
To account for both AGB sub-populations we need to know, for each stellar 
initial mass, the ages marking the transition from (i) oxygen-rich chemistry 
to carbon-rich chemistry, (ii) optically bright source to optically obscured 
source, and (iii) TP-AGB phase to Post-AGB phase.
To estimate the age at which an AGB stars becomes obscured, we assume that 
the stars are not longer observable when the optical depth at $1\mu$m 
($\tau_{1\mu m}\propto\,\dot{M}\Psi$, where $\tau_{\lambda}$ is the optical 
depth, $\dot{M}$ the mass-loss rate, and $\Psi$ the dust-to-gas ratio) is 
higher than a critical value ($\tau_{1\mu m}\ga\,\tau_{crit}$). 
See Sec.\ref{dust_env} for more informations about the envelope models used 
to estimate the optical depth.

\begin{figure}
\includegraphics[clip=,angle=0,width=9.cm,height=9.cm]{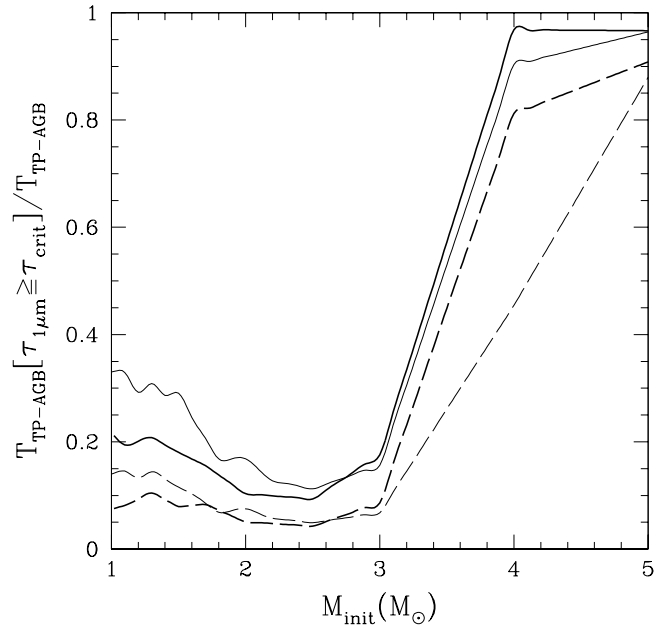}
\caption{Fraction of the total lifetime of the TP-AGB stars spent as
optically obscured stars for solar (thin lines) and LMC (thick line) 
metallicities for $\tau_{1\mu m}=1$ (continuous line) and $\tau_{1\mu m}=3$ 
(dashed line). See text for more details. The mass loss prescription from 
Bl\"ocker (1993) was used with $\eta_B\,\sim\,0.1$.}
\label{TauD_calib}
\end{figure}

Fig.\,\ref{TauD_calib} shows the evolution of the fraction of the whole 
TP-AGB lifetime stars spend as obscured objects, as a function of their 
initial mass for Z=0.02 and Z=0.008 metallicities, for two different values 
of the critical optical depth at $1\mu$m ($\tau_{1\mu m}=1 and 3$).
This choice is justified by the fact that the ingredients (grain properties,
mass-loss rate determination, condensation temperature, ... etc) used to 
construct the envelope models are known within a factor of few (3-5).
The plot shows clearly that massive stars 
(i.e., M$_{init}\,\ge\,4-3.5\,$M$_{\odot}$) that suffer from envelope burning,
spend a large fraction of their TP-AGB lifetime as obscured objects, due to 
their high mass-loss rate (see below).
On the other hand, the plot shows that this fraction decreases rapidly for 
stars having initial masses below this limit. This plot shows again the 
effect of the envelope burning on intermediate age stellar population 
properties, and emphasizes the importance of its inclusion in stellar 
population models.  

\begin{figure}
\includegraphics[clip=,angle=0,width=9.cm,height=9.cm]{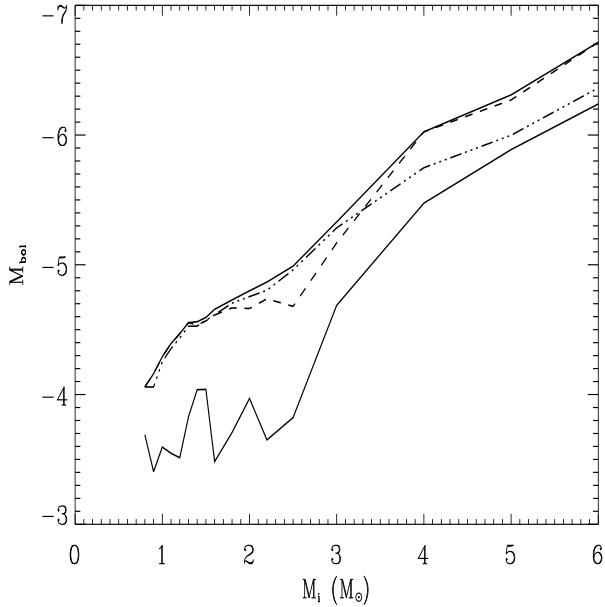}
\caption[]{Bolometric magnitude of important transitions for TP-AGB stars 
with initial metallicity Z=0.02. Shown are the transitions from the E-AGB 
to the TP-AGB (bottom continuous line), the transition from spectral type 
M to spectral type C (dashed line), the end of the TP-AGB (upper continuous 
line), as function of the initial mass. The dotted-dashed line shows the 
evolution of the transition bolometric magnitude for a star from being an 
optically visible to an IR source. We assumed that the star becomes no 
longer optically visible when the optical depth at $1\mu$m is larger 
than 3 (see the text for more information).}
\label{mag_sptype_zsol}
\end{figure}

Figure \ref{mag_sptype_zsol} shows the evolution of the bolometric magnitude 
as a function of the initial mass for the three characteristic stages of the 
AGB phase. The lower and the upper solid lines represent the magnitudes when 
a star enter and quit the TP-AGB phase respectively, the dashed line represents 
the transition magnitude of a TP-AGB star from oxygen-rich to carbon-rich 
chemistry, while the dashed-dotted line represents the transition magnitude 
of a stars from being optically bright to being dust-enshrouded. 
The bolometric luminosity distribution as a function of mass shows that when 
both oxygen-rich and carbon stars coexist in a single age stellar population,
carbon stars are the brightest, as they are more evolved objects than 
oxygen-rich stars. However, for single stellar populations with a turn-off 
mass of M$_{TO}\,\simeq\,4\,M_{\odot}$ (i.e., stellar population age about 
$\sim\,0.2\,$Gyr), for which the competition between the envelope burning 
and the third dredge-up starts to vanish and the latter process becomes more 
effective than in more massive stars, oxygen-rich stars could be brighter. 
Note that at that age, most of the TP-AGB stars are losing mass at a relatively 
large rate. 

Note that massive/luminous carbon stars 
(i.e., M$_{init}\,\ge\,4-3.5\,$M$_{\odot}$) spend their entire lifetime 
as optically invisible object. 
The TP-AGB evolution of those stars is controlled mainly by the interplay 
between the envelope burning, the mass-loss, and the third derdge-up.  
The overluminosity produced by the envelope burning, trigger the super-wind 
($\dot{M}\sim\,10^{-6}-10^{-4}$) very early, and the hot temperature at the 
base of the envelope prevent the formation of carbon stars.
This scheme is valid as long as the envelope mass is high enough to maintain 
a hot temperature at the base of the envelope. When the mass-loss has 
dramatically reduce the mass of the envelope, lowering the temperature 
at the base of the envelope, the envelope burning stops and the star may 
become carbon-rich, but being already embedded in dense circumstellar
shell.

%%%%%%%%%%%%%%%%%%%%%%%%%%%%%%%%%%%%%%%%%%%%%%%%%%%%%%%
The bolometric luminosity distribution of dust-enshrouded M-type stars is 
expected to peak at brighter luminosities than the one of dust-enshrouded
carbon stars, which, in turn, is expected to peak at higher luminosities 
than the luminosity distributions of both optically bright carbon stars 
and M-type stars respectively.
Optically bright carbon stars are expected to populate the AGB mostly between 
M$_{bol}\,\simeq\,-4.3$ and $-5.5$, with a peak around M$_{bol}\,\simeq\,-4.8$,
whilst dust-enshrouded carbon stars may be as bright as M$_{bol}\,\simeq\,-6.5$.
Nevertheless, independently of the initial metallicity dust-enshrouded carbon 
stars as faint as M$_{bol}\,\simeq\,-4.5$ may be found (with progenitor 
masses about M$_{init}\sim\,1.5\,M_{\odot}$).
Note that the peak of the bolometric luminosity distribution of optically bright 
carbon stars is not expected to vary with the metallicity, because the preferred
mass range for carbon star formation is relatively insensitive to metallicity
(for more details, see Mouhcine \& Lan\c{c}on 2002b). 
The luminosity distribution of dust-enshrouded M-type stars is expected to 
have a peak at M$_{bol}\,\simeq\,-6.5$ to $-6.8$. These stars are predicted 
to live $\sim\,4-9\,\times\,10^{4}$ yr in agreement with van Loon et al. (1999).

Note that Fig.\,\ref{mag_sptype_zsol} is constructed neglecting the luminosity 
modulation due to the recurrence of third dredge-up events. This point will be
discussed in more detail in Sec.\,\ref{resol_pop}.  

Olivier et al. (2001) have used a sample of oxygen rich and carbon rich 
dust-enshrouded AGB stars in the solar neighbourhood to estimate empirically 
the duration of the high mass-loss rate AGB phase. 
They show that the stars in their sample have a main-sequence progenitor with 
$1\,$M$_{\odot}\le\,$M$\le\,2\,$M$_{\odot}$ (with an average 
progenitor mass of $\sim\,1.3\,$ M$_{\odot}$) have an average dust-enshrouded 
AGB phase of ($3.7\,\pm\,1.9$)$\times\,10^{4}$ years.
For the same parameters (i.e. M$_{init}\,\approx\,1.3\,$M$_{\odot}$ and 
Z=Z$_{\odot}$), our models predict the high mass-loss AGB phase have a duration 
of $5.5\,\times\,10^{4}$ years, in agreement with the lifetimes derived by 
Olivier et al (2001) (see Habing 1996 for a discussion of other determination 
of the duration of the mass-loss phase).

\subsection{Dust envelope models}
\label{dust_env} 
Extensive work was done to study the properties of circumstellar dust shells 
around late-type stars taking into account radiative transfer and chemistry 
(Hummer \& Rybicki 1971, Bedijn et al. 1978, Rowan-Robinson \& Harris 1983, 
Martin \& Rogers 1984, Justtanont \& Tielens 1992).
However, as our principal goal is to study the dust-enshrouded stellar 
populations as a whole, we need a dust envelope model, which is consistent
with the observational constraints, and makes us able to figure out
(i) the extinction that affects the stellar spectrum due to the dust shell, 
(ii) how much light is emitted in the near-infrared as function of the 
fundamental stellar parameters, and (iii) accounts for the average properties 
of the dust-enshrouded stellar populations.

To construct dusty envelope models around late type AGB stars, the following
assumptions are made: (i) spherical symmetry, and (ii) constant properties 
of the dust grains at different ages and metallicities. We stress that our 
purpose is to have a good description of the average properties of grains in 
circumstellar envelopes. For a given mass loss rate and from the equation 
of continuity, the dust density outside the condensation radius, where the 
dust forms, scales as $r^{-2}$:
\begin{equation}
\rho_{d}(r)\,=\,\frac{\dot{M_{d}}}{4\pi\,r^{2}\,v_{d}}
\end{equation}
where $\dot{M_{d}}$ is the dust mass-loss rate, and v$_{d}$ is the dust 
outflow velocity. The grain emission being negligible in the near-infrared, 
the wavelength range of interest in this paper, the spectral energy 
distribution of mass-losing star is primarily determined by the dust 
optical depth. 
Note that for some extreme stars, grain emission may contributes to a 
relatively large fraction, 20\%-50\%, of K-band light. However these objects 
have very short lifetime and consequently have small effects on stellar 
population properties: a very small number of such stars may present in 
a stellar population, and contribute to a small fraction of integrated 
light budget.\\
The dust optical depth is defined as:
\begin{eqnarray}
\tau_{\lambda}&=&\int^{R_{out}}_{R_c}\pi\,a^2\,
                    Q_{ext}(\lambda)\,n_{d}(r)\,dr \nonumber \\
            && \approx\,\frac{3}{16\pi}\frac{\dot{M}\,\Psi}{R_c\,v_{d}}
                      \frac{Q_{ext}(\lambda,a)/a}{\rho_{gr}} 
\label{opt_depth}
\end{eqnarray}
where Q$_{ext}(\lambda$)=Q$_{abs}(\lambda)$+Q$_{sca}(\lambda)$ is the dust 
extinction, absorption and scattering coefficients respectively, n$_{d}$ is 
the dust grain number density, $a$ is the grain size, $\dot{M}$ is the total 
mass-loss rate, $\Psi$ the dust-to gas 
ratio, $\rho_{gr}$ is the grain density, R$_{out}$ is the outer radius of 
the circumstellar shell, and R$_c$ the dust condensation radius. 
We will assume that the dust velocity equals the terminal velocity of the 
gas. Groenewegen (1993) has shown that this approximation is correct to 
within 10\%.  To derive Eq.\,(\ref{opt_depth}) we have neglected 1/R$_{out}$ 
with respect to 1/R$_{c}$. 

For the near--infrared region we have assumed that the absorption and 
scattering efficiencies of the grains follow a power-law, 
Q($\lambda$)\,=\,Q($1\mu$m)$(\lambda/\mu$m$)^{-\beta}$ (Jura 1983, Bedijn 1987). 
The grain size of both carbon rich dust and oxygen-rich dust was taken to be 
a\,=\,0.1\,$\mu$m, and the grain density was taken to be 
$\rho_{gr}\,=\,2.5\,$gr\,cm$^{-3}$ for silicates and 
$\rho_{gr}\,=\,2.26\,$ gr\,cm$^{-3}$ for amorphous carbon grains. 
The absorption efficiencies at the reference wavelength of 1\,$\mu$m, 
$Q(1\mu$m), were taken from Jones \& Merrill (1976) for silicates and from 
Rowan-Robinson (1986) for amorphous carbon grains. The value of $\beta$ which
fits the energy distribution of carbon stars in the region of our interest is 
$\beta\,=\,1.0$ (e.g. Rouleau \& Martin 1991, see also Rengarajan et al. 1985, 
Le Bertre et al. 1995, Suh 2000). The same value was taken for silicate grains 
(Schutte \& Tielens 1989). We consider silicate grains when dealing with oxygen 
rich circumstellar envelopes, and amorphous carbon grains when dealing with 
carbon rich circumstellar envelopes. We assumed that the properties of all the 
grains in the dusty envelope around a TP-AGB star change instantaneously when 
the star becomes carbon rich. The condensation radius was taken to vary like:
\begin{equation}
\frac{R_c}{R_s}\,\approx\,\frac{1}{2}
                  \left[\frac{T_{eff}}{T{_c}}\right]^{(4+\beta)/2}
\label{Rcond}
\end{equation}
where T${_c}$ is the condensation temperature at the inner radius and R$_{s}$ 
is the stellar radius (i.e., $R_{s}\,\sim\,L^{1/2}\,T_{eff}^{-2}$). 
Groenewegen \& de Jong (1994) have shown, using detailed radiative transfer 
calculations, that the evolution of the condensation radius as function of 
the stellar parameters, at least for carbon stars, is well represented by 
Eq.\,(\ref{Rcond}).  
The condensation temperature of both carbon-rich and oxygen-rich dust is 
taken to be $\simeq\,1000\,$K (Le Bertre 1986, Le Sidaner \& Le Bertre 1996).

To achieve our envelope models, we need to evaluate the dust-to-gas ratio 
$\Psi$ and the expansion velocity $v_{exp}$ for stars evolving 
along the TP-AGB phase. To evaluate $v_{exp}$, 
the period-mass-radius relation of fundamental mode pulsators from 
Wood (1990) is used in combination with the expansion velocity-period 
relation from Vassiliadis \& Wood (1993). Regarding $\Psi$ scaling arguments 
show that $\Psi\,\sim\,v_{exp}^{2}L^{-1/2}$. Using detailed modelling 
Habing et al. (1994) have shown that the exact dependence of $\Psi$ on 
luminosity and v$_{exp}$ deviates slightly from the scaling formula. We will 
use the formula of Habing et al. Note that we do not introduce any additional 
explicit dependence on metallicity in our semi-analytical envelope model.

\section{Effects of obscured AGB stars.}
\label{pro}
In the following subsections, we will discuss how dust-enshrouded AGB stars
affect near-infrared properties of both resolved galaxies and unresolved 
stellar populations. For this purpose, we coupled the dusty envelope model 
with synthetic AGB models. Regarding resolved populations, we study 
the sensitivity of the AGB sequence morphology in the near-infrared HR 
diagram to the dominant AGB subtype. For unresolved stellar populations 
we will examine how 
optical/near-infrared broad-band colours are sensitive to the presence 
of obscured AGB stars.

Transformations from the theoretical to the observational plane were 
performed by making use of the stellar spectral library of 
Lan\c{c}on \& Mouhcine (2002). The latter library is based on stars on 
the solar neighbourhood. To account for a dependence on the metal content 
even for M giants, we adopted the same library at different metallicities 
but assigned the spectral class, identified by the (I-K) colour, adopting 
the (I-K)-T$_{eff}$ relation of Bessell et al. (1989) which depends on the 
metallicity. (R-H) was taken to be an effective temperature indicator of 
carbon stars (Loidl et al. 2001). The sensitivity of carbon star spectra 
to metallicity is thought to be small (Gautschy 2001). Hence carbon star 
spectra of solar neighbourhood stars were used at all metallicities.  
  
To account for the effect of the dust along the TP-AGB, we combine the 
AGB star synthetic evolution models and the semi-analytical envelope models 
presented in Sec.\,\ref{dust_env}. The optical depth is a combination of 
two effects. The first one (i.e., Q($\lambda$)/a\,$\rho_{gr}$) describes 
the circumstellar shell chemistry 
and the second one describes the effect of the evolutionary status of a 
TP-AGB star on its circumstellar shell structure. At fixed age, we use 
stellar evolutionary parameters of the TP-AGB stellar population that 
populate the isochrone (i.e., T$_{eff}$, L, M$_{init}$, Z/Z$_{\odot}$) 
to derive the optical depth for each star, and hence to compute the 
near-infrared spectral energy distribution of mass-losing stars 
(Volk \& Kwok 1988, Bressan et al 1998).

\subsection{Resolved stellar populations: Isochrones in the Infrared }
\label{resol_pop}
\begin{figure*}
\includegraphics[clip=,angle=0,width=9.cm]{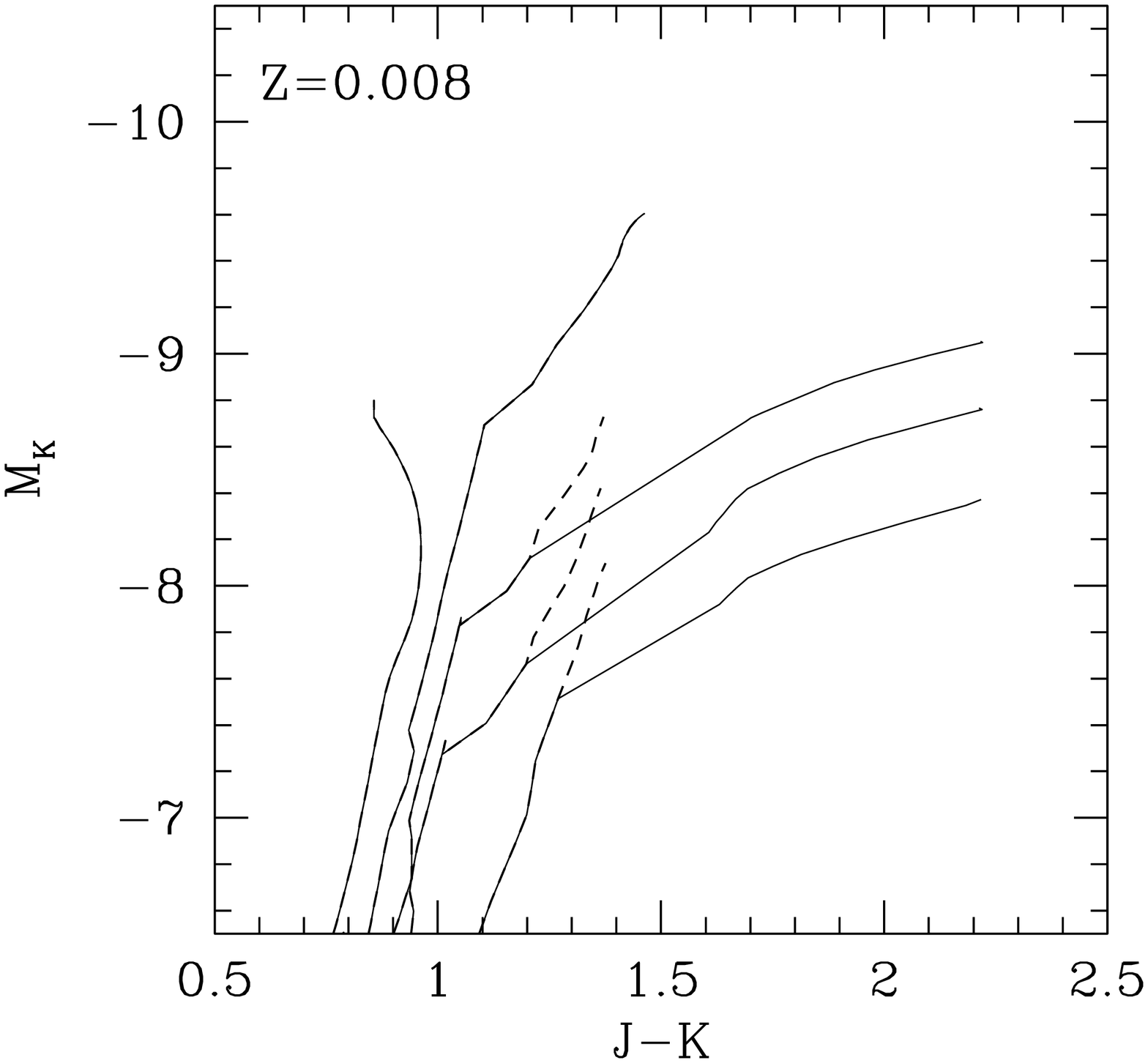}
\includegraphics[clip=,angle=0,width=9.cm]{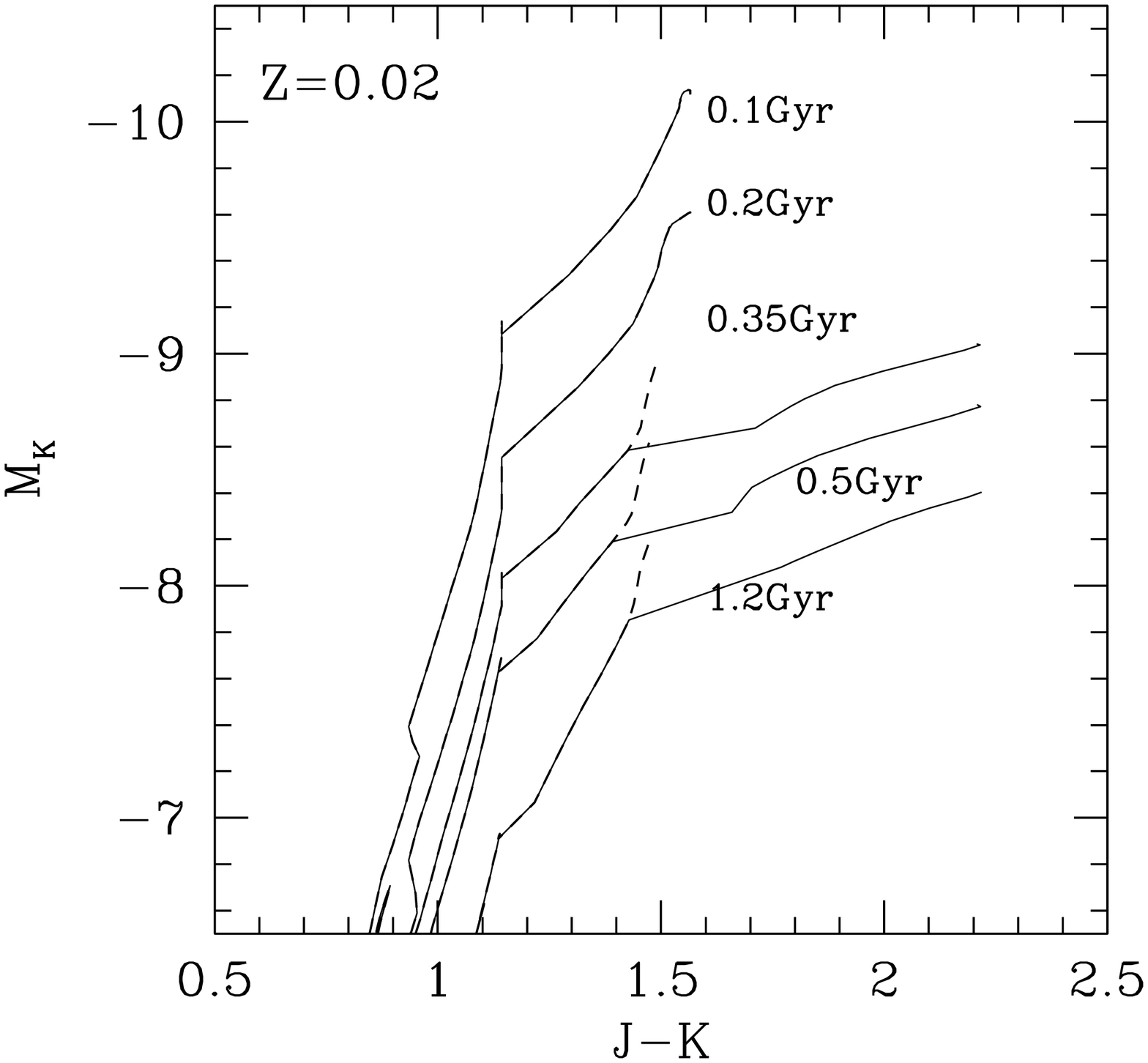}
\caption[]{Synthetic near-infrared colour-magnitude diagram of selected
isochrones with Z=0.008 and Z=0.02 where the formation of the dust shells 
is ignored. Continuous lines show isochrones constructed when both oxygen-rich 
stars and carbon stars are considered, while the dashed lines represent the 
isochrones when carbon star formation is ignored. }
\label{mk_jmk_op}
\end{figure*}

First, we examine the effect of carbon stars on the morphology of the upper 
cool/luminous part of the colour-magnitude diagram without considering 
obscuration. Fig\,\ref{mk_jmk_op} shows selected isochrones with Z=0.008 
and Z=0.02 in the M$_{K}$ vs. (J-K) diagram.

\begin{figure*}
\includegraphics[clip=,angle=0,width=9.cm]{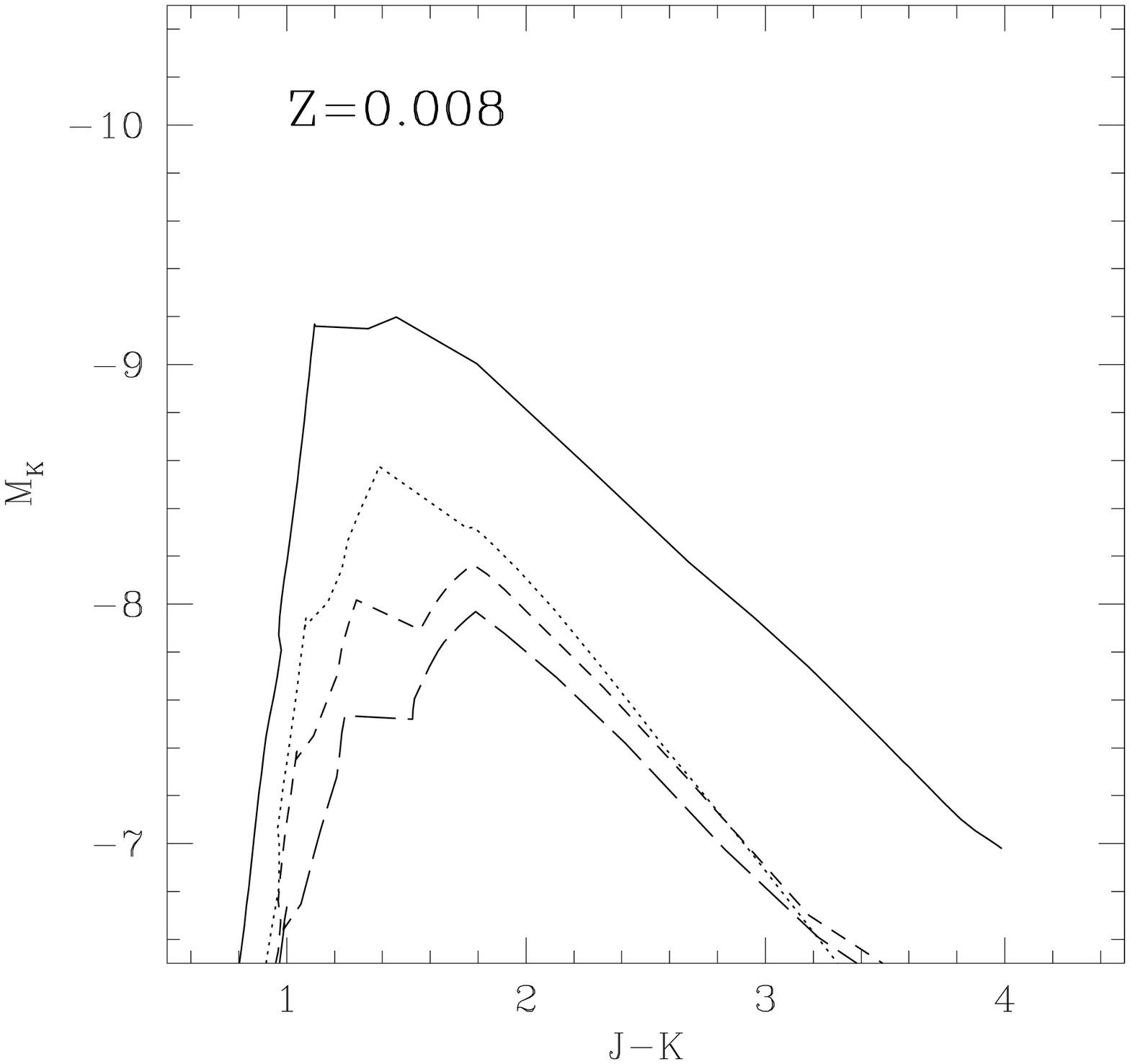}
\includegraphics[clip=,angle=0,width=9.cm]{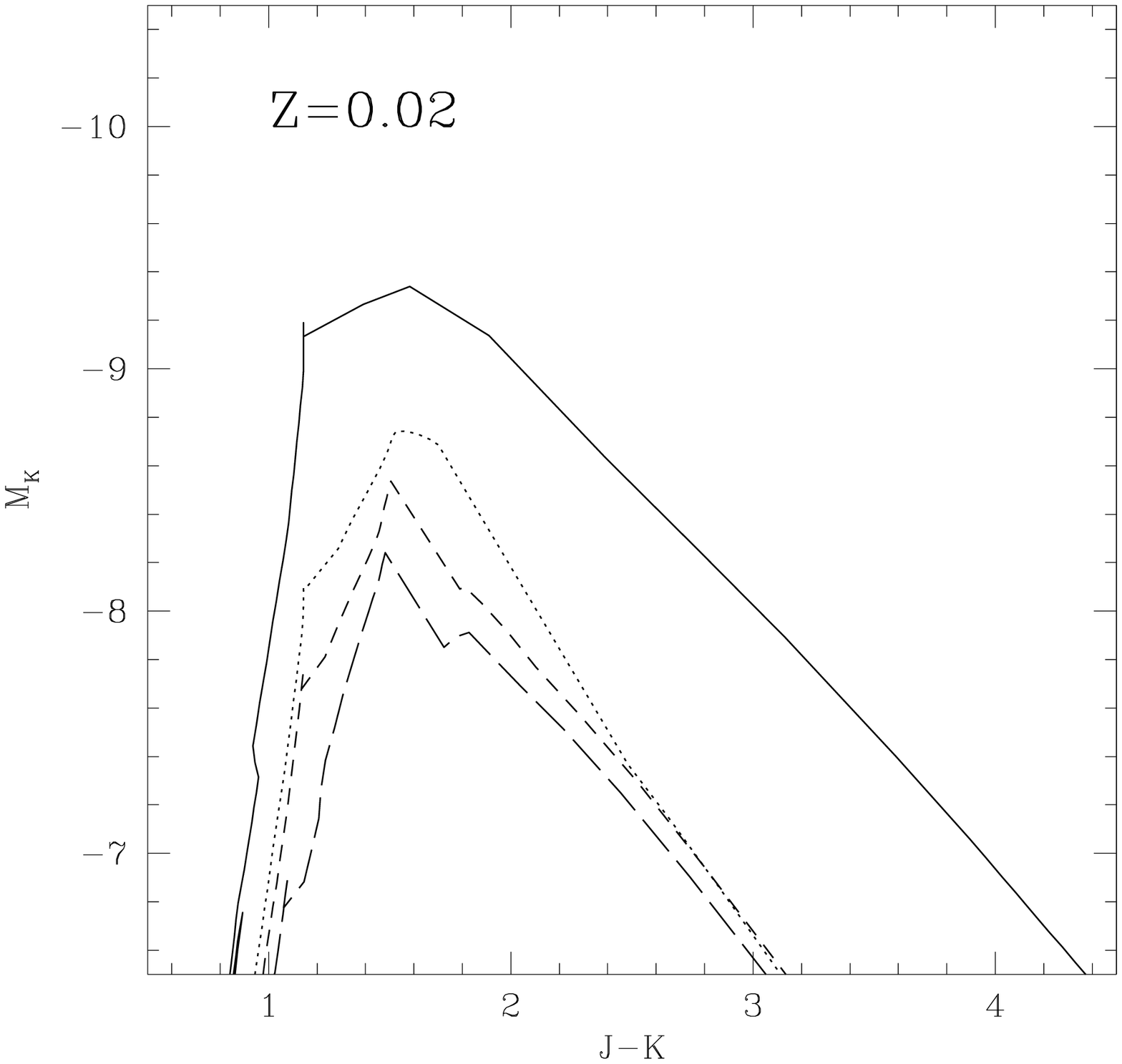}
\caption{Synthetic near-infrared colour-magnitude diagram of 
selected isochrones with Z=0.008 and Z=0.02 (continuous line: 
0.12\,Gyr, dotted line: 0.35\,Gyr, dashed line: 0.5\,Gyr, and 
long-dashed line: 0.8\,Gyr). Both the formation of carbon stars 
and the effects of dust envelope on the spectral properties are 
taken into account. }
\label{col_mag}
\end{figure*}

\begin{figure*}
\includegraphics[clip=,angle=0,width=9.cm]{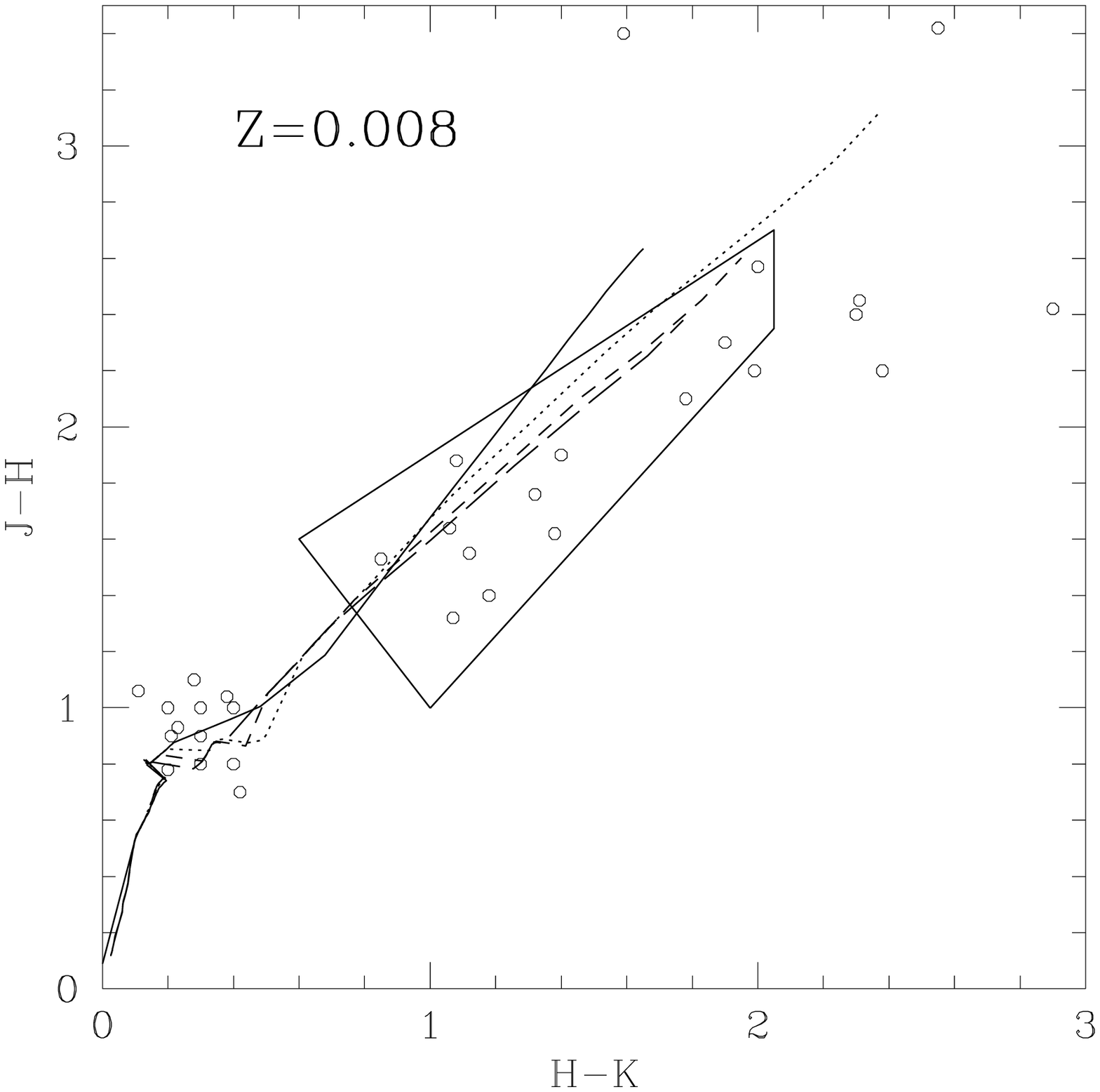}
\includegraphics[clip=,angle=0,width=9.cm]{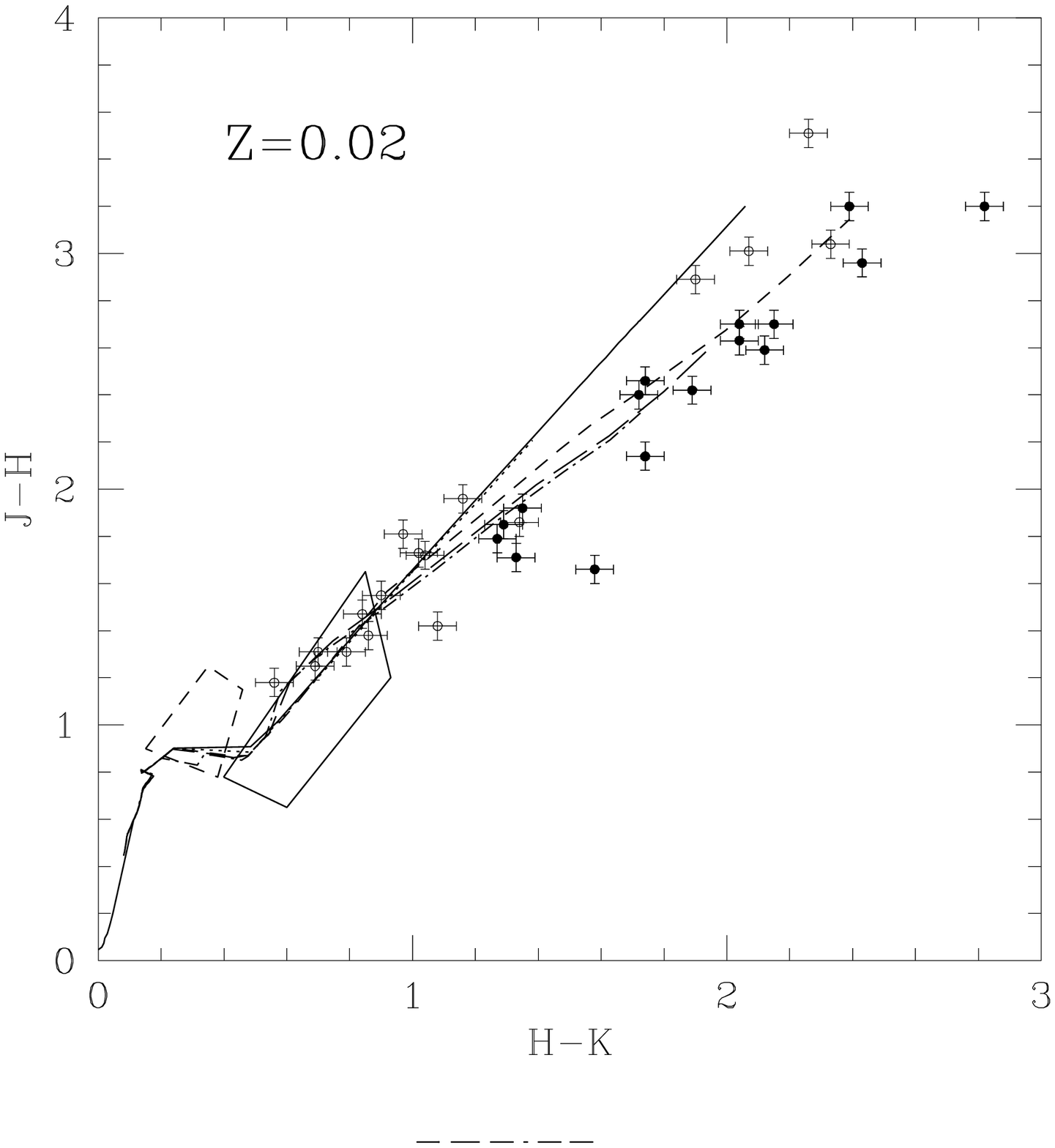}
\caption{H-K vs. J-H synthetic two colour diagram. The upper panel 
shows some selected isochrones with metallicity Z=0.008 (continuous 
line: 0.12\,Gyr, dotted line: 0.2\,Gyr, dashed line: 0.35\,Gyr, 
long-dashed line: 0.5\,Gyr and dotted-dashed line: 0.8\,Gyr). 
Overplotted is the sample of obscured AGB stars in the LMC from 
Zijlstra et al (1996). The region outlined 
by a solid line encompasses LMC sources with J-K$\,>\,2$ as observed 
by 2MASS (Nikolaev \& Weinberg, 2000). The lower panel shows some 
selected isochrones with metallicity Z=0.02. Overplotted are Olivier 
et al.'s (2001) sample of dust-enshrouded object in the solar 
neighbourhood.
Open circles represent oxygen rich stars, filled circles carbon stars. 
The regions outlined by the solid line and the dashed line encompass 
South Galactic Cap Miras stars and M stars (Whitelock et al. 1994, 
1995) respectively.}
\label{col_col}
\end{figure*}

%%%%%%%%%%%%%%%%%%%%%%%%%%%%%%%%%%%%%%%%%%%%%%%
Continuous lines show isochrones where oxygen-rich 
and carbon stars are accounted for. Dashed lines show isochrones for which 
only oxygen-rich stars are considered. The plot illustrates nicely the 
effect of carbon stars on the morphology of the upper colour-magnitude 
diagram of intermediate-age stellar populations. When carbon stars are 
neglected, the models predict that AGB star populations 
should appear as a single and continuous "plume" in the M$_{K}$ vs. (J-K) 
diagram, with (J-K)$\,\sim\,0.8-1.5$. However, when carbon star formation 
is followed, a {\it{carbon star branch}} is expected. This branch will 
appear in the 
colour-magnitude diagram as a second plume where (J-K) may be as red as 
2 or redder for the same range of K-band magnitudes than the models that 
neglect the formation of carbon stars. The models predict that the first 
plume is occupied 
by oxygen-rich stars. Isochrones younger than $\sim\,0.2-0.3\,$Gyr 
are not affected by the inclusion of carbon stars in the modelling for 
the simple reason that carbon stars are not present yet in the stellar 
population. The models predict that the {\it{carbon stars branch}} in 
the colour-magnitude diagram is more pronounced at Z=0.008 than at Z=0.02, 
because of the higher efficiency of carbon star formation in metal-poor 
systems. 

%%%%%%%%%%%%%%%%%%%%%%%%%%%%%%%%%%%%%%%%%%%%%%%
Fig.\,\ref{col_mag} shows selected isochrones with Z=0.008 and Z=0.02 
in the M$_{K}$ vs. (J-K) diagram. Now the formation of both carbon stars
and circumstellar shells around the TP-AGB stars are accounted for to 
construct isochrones.
A new morphological feature appears in the colour-magnitude diagram as 
an additional plume. The plot shows the location of three different 
populations in the M$_{K}$ vs. (J-K) diagram. The first stellar 
sub-population have $0.6\,\la\,(J-K)\,\la\,1.4$. This region of the HR 
diagram is occupied by Early-AGB stars and oxygen-rich TP-AGB stars. 
The magnitude extension of this sub-population is quite large
because it is populated by stars with different progenitor initial mass 
(from 5-6\,M$_{\odot}$ to $\sim\,1\,$M$_{\odot}$), with quite different 
magnitude at the tip of the AGB. The second stellar sub-population have 
$1.4\,\la\,(J-K)\,\la\,2$, and it contains optically bright carbon-rich 
TP-AGB stars. The luminosity extent of this population is significantly 
narrower than that of the first population. The extent of the morphological 
feature related to this population is less pronounced at solar metallicity 
than at Z=0.008 metallicity, because of the larger efficiency of carbon 
star formation at low metallicity.  
The third stellar sub-population has extremely red colours
(i.e., J-K$\,\ga\,2$), and decreasing K-band luminosity as the colour 
becomes redder. This region contains extremely red dust-enshrouded TP-AGB  
stars, both oxygen-rich and carbon-rich. Their large (J-K) and decreasing 
K-band luminosity are a consequence of large extinction due to dusty 
circumstellar envelopes.

The predicted morphology of this region of the HR diagram is in good 
agreement with the HR diagram morphology of the LMC stellar populations 
as observed by 2MASS (Nikolaev \& Weinberg 2000).
%%%%%%%%%%%%%%%%%%%%%%%%%%%%%%%%%%%%%%%%%%%%%%%

To check further the behavior of our models in the near infrared region 
of the spectrum, we have compared our isochrones with samples of optically
bright and dust-enshrouded stars. 
Fig.\,\ref{col_col} (right panel) shows the selected isochrones at solar 
metallicity in the two colour diagram (J-H) vs. (H-K). Also shown is 
the location of M giants (area delineated by dashed line) and Miras 
(area delineated by continuous line) as observed in the Southern Polar 
Cap (Whitelock et al. 1994, 1995), in addition to carbon (filled circles) 
and oxygen-rich (open circles) dust-enshrouded AGB stars in the solar 
neighbourhood (Olivier et al. 2001). The left panel shows the same 
isochrones at Z=0.008 metallicity, compared to a sample of 
dust-enshrouded AGB stars in the LMC (Zijlstra et al. 1996), and the 
location of LMC/2MASS sources with (J-K)$\,\geq\,2$ 
(Nikolaev \& Weinberg 2000).

Both the bulk of the M giants at (J-H)$\,\simeq\,1$ and (H-K)$\,\simeq\,0.3$ 
and of Miras samples are well reproduced by the models. 
As the star reaches the TP-AGB phase, the effect of the mass-loss increases 
and the circumstellar shells modify the energy distributions. Isochrones 
with different ages and metallicities reach different values of extinction. 
In the framework of our semi-analytical envelope model, the optical depth is 
an increasing function of mass and metallicity. The larger 
the metallicity and the lower the isochrone age, the larger the optical 
depth, and the further the isochrone extends in the colour-colour diagram. 
TP-AGB stars losing mass at high rates populate an extended {\it{red arm}} 
in the (J-H) vs. (H-K) two colour diagram. Fig.\,\ref{col_col} shows that 
this {\it{red arm}} splits into two arms occupied by carbon stars and oxygen 
rich stars respectively. Dust-enshrouded carbon stars exhibit different 
colours from dust-enshrouded M-stars, indeed they have a redder (H-K) colour 
than M-type stars at fixed (J-H).
This predicted behavior is in agreement with the observational constraints 
in the solar neighbourhood (Olivier et al. 2001).
This behavior stresses again the importance of the inclusion of both AGB 
subtypes simultaneously to model intermediate-age stellar populations. 

We have to mention that a fraction of carbon star subpopulation may 
occupy bluer and fainter regions in the HR diagram, than what the bulk 
of the subpopulation does. These relatively {\it{blue}} carbon stars may 
be low-mass carbon stars evolving through the faint/hot luminosity-dip 
phase of the quiescent hydrogen burning phase that follows immediately 
a thermal pulse. It is known that this luminosity-dip phase affects 
strongly the faint tail of carbon stars luminosity function (Marigo et 
al. 1996, see also below). If the star formation in a galaxy proceeds 
continuously, one may expect that a fraction of carbon stars should mix 
with oxygen-rich stars at bluer colours; this was shown by Kontizas et 
al. (2001) in their survey of carbon stars in the LMC where they found 
that a significant number of spectroscopically selected carbon stars are 
bluer and fainter than "carbon star's region" defined photometrically by 
Nikolaev \& Weinberg (2000). The implication of the presence and the 
properties of this subpopulations on the the link between carbon star 
statistics and the star formation history, and hence the usage of carbon 
stars as quantitative star formation tracer are left to a forthcoming paper, 
where we present comparisons of our models to data (Mouhcine et al. 2002, 
in preparation).
%%%%%%%%%%%%%%%%%%%%%%%%%%%%%%%%%%%%%%%%%%%%%%%%%%%%%%%%%%%%%%
\begin{figure}
\includegraphics[clip=,angle=0,width=9.cm,height=12.cm]{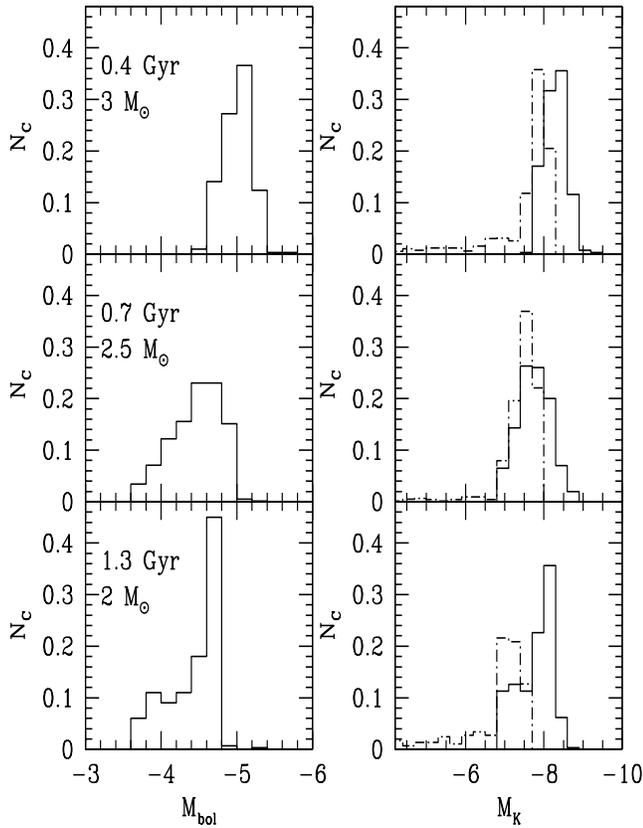}
\caption{Theoretical carbon star bolometric (left panel) and K-band (right
panel) luminosity functions for simple stellar populations of initial
metallicity Z=0.008. Each histogram corresponds to the predicted
distribution of carbon stars of the same age, i.e. evolved from progenitors
with the same initial mass, as indicated. The dashed-dotted line is the
K-band luminosity function of carbon stars when the effect of circumstellar
envelope on the energy distribution is accounted for.}
\label{cslf}
\end{figure}

The results and discussions presented until now were drawn out using 
{\it{smooth}} tracks; i.e. tracks where the stellar luminosity modulation 
by thermal pulses was averaged out and the resulting brightening rate of 
carbon stars is nearly constant. In fact TP-AGB stars undergo a long-lived 
underluminous stage, soon after the occurrence of a thermal pulse, with deeper 
depth and longer duration for low-mass stars (M$_{init}\le\,3\,$M$_{\odot}$). 
This effect plays a crucial role in determining the extension of the 
low-luminosity tail of the luminosity function
of carbon stars (Boothroyd \& Sackmann 1988, Groenewegen \& de Jong 1993).

Fig.\,\ref{cslf} shows bolometric and K-band luminosity functions of carbon 
stars that exist in simple stellar populations of selected ages, where roughly 
all carbon stars have the same initial mass of the progenitor noted in each 
panel. From this figure, it turns out that the faint tail of the bolometric 
luminosity function extends $\sim\,0.8-1$ mag fainter than what is expected 
from the smoothed tracks.  
Stars with initial masses M$_{init}\le\,2.5\,$M$_{\odot}$ have luminosity 
functions that peak almost at the same magnitude (see also Marigo et al. 1999). 
Note that few stars brighter than the tip of the AGB magnitude exist in the 
stellar population. Those stars evolve on the rapid luminosity peaks related 
to the thermal pulses. Additional large-amplitude pulsations of the AGB stars 
could affect the magnitude extent of different AGB sub-populations. Bolometric 
amplitudes of galactic carbon stars are $\sim\,0.6$ mag (Le Bertre 1992) and 
may be similar to those of galactic OH/IR stars (Le Bertre 1993). 

The right panel of Fig.\,\ref{cslf} shows comparisons between K-band luminosity 
functions of carbon stars of populations where all carbon stars are considered 
to be optically bright (continuous line), and stellar populations where both 
optically-bright and dust-enshrouded carbon stars coexist (dashed-dotted line). 
Two features arise. (i) When both sub-types of carbon stars are considered, the 
peak of the luminosity functions shifts to fainter magnitudes in comparison to 
what is expected when the formation of shells around carbon stars is ignored, 
and (ii) a faint and extended luminosity function tail (M$_{K}\ge\,-5$) is 
predicted. The stars that populate this faint wing are mainly stars close to 
the end of their AGB phase having high mass-loss rate and extreme red colours 
(J-K$\,\ge\,3$), and most of their light is emitted at wavelengths longer than 
the K-band. As a consequence of the formation of this faint tail, the main body 
of the K-band luminosity function becomes narrower than what is predicted when 
the formation of dust-enshrouded carbon stars is neglected. 

\subsection{Unresolved stellar populations}

\begin{figure*}[ht]
\includegraphics[clip=,angle=0,width=13.cm,height=11.cm]{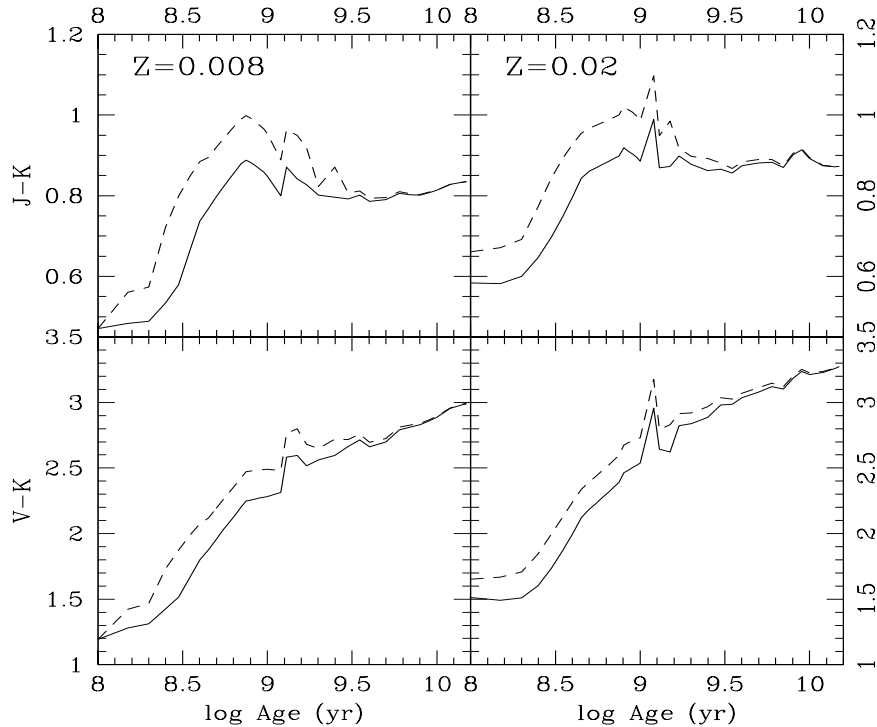}
\caption[]{Temporal evolution of optical and near infrared colours of simple
stellar population at solar and Z=0.008 metallicities. Continuous lines refer
to models constructed taking into account the presence of circumstellar shells
around TP-AGB stars, while dashed-lines refer to ones when these stars were
neglected. }
\label{IR_colours}
\end{figure*}

In this section we will examine the effect of the dusty envelopes around TP-AGB
stars on the integrated near-infrared properties of intermediate-age stellar 
populations. We will restrict these investigations to the case of instantaneous 
burst models (occurring at t=0).

Once the optical depth of a TP-AGB star is calculated, its "reddened" 
spectrum is added to the stellar population spectrum weighted by the initial 
mass function (hereafter IMF). In the rest of the paper we adopt a Salpeter 
IMF between the lower and upper cut-off masses, M$_{low}=\,0.1\,M_{\odot}$ 
and M$_{up}=\,120\,M_{\odot}$ respectively 
($\phi(m)\propto m^{-\alpha}, \alpha\,=\,2.35$).
Fig.\,\ref{IR_colours} shows the temporal evolution of selected optical/near-IR 
colours when the formation of dust enshrouded AGB stars is taken into account 
at solar and Z=0.008 metallicities. For comparison, the evolution of 
optical/near-IR colours neglecting the effects of circumstellar shells is also 
plotted. The most striking feature is that models that account for the formation 
of dust-shells around the TP-AGB stars have bluer colours than models that do 
not account for them (see also Bressan et al. 1998). This behavior is due to 
the large obscuration affecting the brightest TP-AGB stars leading to a decrease 
of their near-IR flux. Keep in mind that the optical light is dominated by 
other stellar populations than AGB stars. Using these models for age-dating of 
stellar populations will lead to assigning older ages than what is predicted by 
models assuming pure photospheric emission.
As the age of stellar population proceeds, the departure from the predictions 
of pure photospheric emission models decreases translating the fact that red 
giant branch stars become the dominant source of the near-IR spectral energy.

\begin{figure}[!ht]
\includegraphics[clip=,angle=0,width=9.cm]{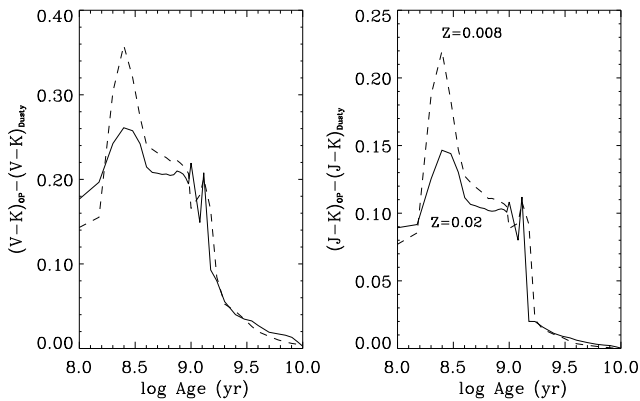}
\caption[]{Differences between the synthetic broad-band colours of models 
that differentiate optically-bright and dust-enshrouded stars (labelled 
"Dusty") and models that neglect the formation of the these stars (labelled 
"OP") for both metallicities indicated in the plots.}
\label{opt_dust}
\end{figure}

Metallicity effects are depicted in Fig.\,\ref{opt_dust} where we plot the 
temporal evolution of differences between the synthetic colours of models that 
differentiate optically-bright and dust-enshrouded stars and models that do not 
for Z=0.02 and Z=0.008.
For young stellar populations (i.e. age$\,\le\,0.25\,$Gyr), when carbon stars 
do not yet exist, higher metallicity models are more affected by the inclusion 
of dust shells around TP-AGB stars because of their low effective temperatures 
which lead to higher mass loss rates and hence higher optical depth. 
The bump-like shape that appears around $\,\sim\,0.25\,$Gyr and extends until 
$\,\sim\,1\,$Gyr is related to the presence of carbon stars in the stellar 
populations. During this age range the effect of dust shells is larger at 
low metallicity even though metal-poor stars have lower mass loss rates than 
metal-rich ones.
This is explained by the fact that at lower metallicity the fraction of carbon 
stars is higher, and that the extinction coefficients of carbonaceous grains 
found around carbon stars are larger than those of silicate grains surrounding 
oxygen-rich stars. The latter dominate the light of the stellar populations 
at the same age interval at high metallicity. 
In addition, AGB stars become brighter as the stellar metallicity decreases, 
increasing the optical depth. Hence carbon stars are affected more severely 
than oxygen-rich stars having the same location location in the HR diagram 
leading to a bigger departure from the prediction of pure photospheric emission 
models. For older stellar populations, as the contribution of AGB stars goes 
down, no envelope-related metallicity effect is seen.

%-----------------
\section{Conclusions}
\label{con}
In this paper we have modelled the effects of both carbon stars and of 
the presence of a steady-state outflow of matter surrounding TP-AGB stars
on the optical and near-IR emission of  
intermediate-age stellar populations. To this aim we coupled stellar 
evolution models and dusty shell models, in order to derive the evolution 
of the extinction due to circumstellar shells along 
the TP-AGB phase. Effects on the luminous and cool region of the HR 
diagram of resolved stellar 
populations and on integrated near-infrared properties of unresolved 
intermediate-age stellar populations were investigated. 

The effects of both AGB sub-populations are predicted to be strong on the 
distribution of AGB stars in the colour-magnitude diagram. Indeed, the 
AGB sequence is expected to split into three morphological features 
occupied respectively by oxygen-rich Early-AGB/TP-AGB stars, carbon stars, 
and dust-enshrouded stars with little overlap, in good agreement with the 
feature-rich 2MASS LMC colour-magnitude diagram. 

As a test of our models, we compared our isochrones with a sample of 
galactic M stars, Mira stars, and extreme dust-enshrouded objects in the 
Galaxy and in the LMC. In the near infrared colours the models reproduce 
the location of both optically bright and dust-enshrouded stars in a 
(J-H) vs. (H-K) diagram. The models predict that the dust-enshrouded carbon 
star population populates a sequence with redder (H-K) than dust-enshrouded 
M-star population at fixed (J-H).

We show that the K-band luminosity function of carbon stars is sensitive to 
the inclusion of the effect of dust-shells around TP-AGB stars. The peak of 
the luminosity function is expected to shift to fainter magnitudes and faint 
tails populated by cool, luminous but obscured stars close to the end of 
their AGB phase are predicted.

Regarding integrated properties of stellar populations, the inclusion of dust 
shells around AGB stars reduces the contribution of these stars to near-IR 
light because of the large obscuration affecting them. This will lead to bluer 
optical/near-IR colours in comparison to what is predicted by models where 
the formation of circumstellar shells is neglected. This effect decreases as 
the stellar population grows older.

\begin{acknowledgements}
I would like to thank A. Lan\c{c}on, and C. Loup for their insightful remarks
on earlier version of this paper, and the anonymous referee for her/his 
constructive report.
\end{acknowledgements}

%-----------------


\begin{thebibliography}{}
\bibitem{} Aaronson, M., \& Mould J., 1985, ApJ 290, 191
\bibitem{} Bedijn P.J., 1987, A\&A 186, 136
\bibitem{} Bessell, M.S., Brett, J.M., Scholz, M., \& Wood, P.R., 1989, A\&AS 77, 1
\bibitem{} Bl\"ocker, T., 1995, A\&A 297, 727
\bibitem{} Boothroyd A.I., Sackmann I.-J., 1988, ApJ 328, 632
\bibitem{} Bressan, A., Granato, G.L., \& Silva L., 1998, A\&A 332, 135
\bibitem{} Cioni, M.-R.L., Loup, C., Habing, H.J., et al., 2000, A\&AS 144, 235
\bibitem{} Costa, E., Frogel, J.A., 1996, ApJ 112, 2607 
\bibitem{} Ferraro, F., Fusi Pecci, F., Test, V., et al., 1995, MNRAS 272, 391
\bibitem{} Forestini, M., \& Charbonnel, C., A\&A, 1997, A\&AS, 123, 241
\bibitem{} Frogel, J.A., Mould, J.R., Blonco, V.M., 1990, ApJ 352, 96
\bibitem{} Frost C.A., Cannon R.C., Lattanzio J.C., Wood P.R., \& Forestini M.,
           1998, A\&A 332, L17
\bibitem{} Gautschy, R., 2001, PhD thesis, University of Vienna, Austria
\bibitem{} Groenewegen, M.A.T., \& de Jong, T., 1993, A\&A 267, 410
\bibitem{} Habing, H.J., Tignon, J., \& Tielens, A.G.G.M., 1994, A\&A 286, 523 
\bibitem{} Habing, H.J.,  1996, A\&ARv, 7, 97
\bibitem{} Herwig, F., Sch\"onberner, D.,Bl\"ocker, T., 1998, A\&A, 340, 43
\bibitem{} Herwig, F., 2000, A\&A, 360, 952
\bibitem{} Ivezic, Z., \& Elitzur M., 1995, ApJ 445, 415
\bibitem{} Jones T.W., \& Merrill K.M., 1976 ApJ 209, 509
\bibitem{} Jura, M., 1983, ApJ 267, 647
\bibitem{} Justtanont, K., \& Tielens, A.G.G.M., 1992, ApJ 389, 400
\bibitem{} Kontizas, E., Dapergolas, A., Morgan, D.H., Kontizas, M., A\&A, 2001, 369, 932
\bibitem{} Lamers, H.J.G.L.M., \& Cassinelli, J.P., {\it{Introduction to stellar 
           winds}}, Cambridge University Press, 1999
\bibitem{} Lan\c{c}on, A., \& Mouhcine, M., 2002, A\&A, in press	   
\bibitem{} Le Bertre, T., 1989, A\&A 203, 85
\bibitem{} Le Bertre, T., 1992, A\&AS 94, 377 
\bibitem{} Le Bertre, T., 1993, A\&AS 97, 729
\bibitem{} Le Sidaner, P., \& Le Bertre, T., 1996, A\&A 314, 896
\bibitem{} Loidl, R., Lan\c{c}on, A., \& Jorgensen, U.G., 2001, A\&A 371, 1065
\bibitem{} Marigo, P., 1998, A\&A 340, 463
\bibitem{} Marigo, P., Girardi, L, \& Bressan, A., 1999, A\&A 344, 123
\bibitem{} Marigo, P., Girardi, L., Weiss, A., \& Groenewegen, M.A.T. 1999, A\&A 351, 161
\bibitem{} Martin, P.G., \& Rogers, C., 1987, ApJ 322, 374
\bibitem{} Mouhcine, M., \& Lan\c{c}on, A., 2002a, A\&A, accepted
\bibitem{} Mouhcine, M., \& Lan\c{c}on, A., 2002b, submitted
\bibitem{} Mouhcine, M., et al., in preparation
\bibitem{} Omont, A., Ganesh, S., Alard, C., et al., 1999, A\&A 348, 755 
\bibitem{} Olivier, E.A., Whitelock, P.A., Marang, F., 2001, MNRAS, 326, 490 
\bibitem{} Nikolaev, S., \& Weinberg, M.D., 2000, ApJ 542, 804
\bibitem{} Rebeirot, E., Azzopardi, M., \& Westerlund, B.E., 1993, A\&AS 97, 603
\bibitem{} Rowan-Robinson, M., \& Harris S., 1983, MNRAS 202, 767
\bibitem{} Rowan-Robinson, M., 1986, MNRAS 219, 737
\bibitem{} Rouleau, F., \& Martin, P.G., 1991, ApJ 377, 526
\bibitem{} Persson, S. E., Aaronson, M., Cohen, J. G., Frogel, J.A.,
           \& Matthews, K., 1983, ApJ 266, 105
\bibitem{} Skrutskie M., 1998, in Epchtein N., ed., {\it{The Impact of 
           Near-Infrared Sky Surveys on Galactic and Extragalactic Astronomy}}. 
	   Kluwer Academic, Dordrecht, p. 11
\bibitem{} Schutte, W.A., \& Tielens, A.G.G.M., 1989, ApJ 343, 369
\bibitem{} van Loon, J.T., Groenewegen, M.A.T., de Koter, A., et al., 1999, 
           A\&A 351, 559
\bibitem{} Volk, K., \& Kwok, S., 1988,, ApJ 331, 435
\bibitem{} Wagenhuber J., Groenewegen M.A.T., 1998, A\&A, 340, 183
\bibitem{} Whitelock, P., Menzies, J., Feast, M., et al., 1994, MNRAS 267, 71
\bibitem{} Whitelock, P., Menzies, J., Feast, M., et al., 1995, MNRAS 276, 219
\bibitem{} Zaritsky, D., Harris, J., Thompson, I.B., 1997, AJ 114, 1002
\bibitem{} Zijlstra, A.A., Loup, C., Waters, L.B.F.M., et al., 1996, MNRAS 279, 32
\end{thebibliography}
\end{document}